\documentstyle{article}
\title{\strut
\rightline{\normalsize\it                Submitted to Teor. i Mat. Fiz.}
\bigskip\newline
                  Discrete analogues of the Liouville equation}
\author{                V.E. Adler, S.Ya. Startsev}
\date{\null}
\font\Sets=msbm10 \def\Z{\hbox{\Sets Z}}
\def\phi{\varphi}
\def\eps{\varepsilon}

\def\pa {\partial}
\def\ord  {\mathop{\rm ord}  \nolimits}

\def\be  {\begin{equation}}    \def\ee  {\end{equation}}
\def\ba  {\begin{array}}       \def\ea  {\end{array}}
\def\bea {\begin{eqnarray}}    \def\eea {\end{eqnarray}}
\def\bean{\begin{eqnarray*}}   \def\eean{\end{eqnarray*}}
\newtheorem{theorem}{Theorem}
\newtheorem{lemma}[theorem]{Lemma}
\def\proof{\paragraph{Proof.}}
\def\qed{\vrule height0.6em width0.3em depth0pt}
\begin{document}
\maketitle

\begin{abstract}
The notion of Laplace invariants is transferred to the lattices and discrete
equations which are difference analogs of hyperbolic PDE's with two
independent variables.  The sequence of Laplace invariants satisfy the
discrete analog of twodimensional Toda lattice.  The terminating of this
sequence by zeroes is proved to be the necessary condition for existence of
the integrals of the equation under consideration.  The formulae are
presented for the higher symmetries of the equations possessing integrals.
The general theory is illustrated by examples of difference analogs of
Liouville equation.
\end{abstract}

\section{Introduction}

Liouville equation and its generalizations are studied about one hundred
fifty years and very many results, including classification ones, are
obtained in this field \cite{Lio} -- \cite{JuA}.  Surprisingly, it seems that
in the discrete version these investigations were not undertaken, although
many results can be transferred from the continuous case without any efforts.
The aim of our paper is to advance in this direction.  The equations with
explicit dependence on the discrete variable are most difficult and for the
present we consider only shift-invariant equations although this condition
is, probably, too restrictive.  Overcoming of this difficulty as well as
deduction of classification theorems require further researches.

Let us consider differential-difference and totally discrete equations of the
form
\be                                                         \label{uix}
    u_{i+1,x}=f(x,u_{i+1},u_i,u_{i,x}),
\ee
\be                                                         \label{uij}
    u_{i+1,j+1}=f(u_{i+1,j},u_{i,j+1},u_{i,j}).
\ee
Further on we assume that for the equation (\ref{uix}) $f_{u_{i,x}}\ne0$, and
the right hand side of the equation (\ref{uij}) depends explicitly on each of
the indicated variables.  Apparently these equations can be treated as
difference approximations of the equations of the form
\be                                                         \label{uxy}
    u_{xy}=f(x,y,u,u_x,u_y).
\ee
They arise also as B\"acklund transformations and nonlinear superposition
principles, and of course this concern both generalizations of the Liouville
equation and equations integrable by Inverse Scattering Transform (the well
known examples are dressing chain \cite{VesSh} and difference KdV equation
\cite{PNC}).

One can divide the different generalizations of the Liouville equation into
several (intersecting) subclasses, subject to what property of the Liouville
equation is chosen as definition of integrability, for example: explicit
formula for the general solution, differential substitution to linear
equation, symmetries depending on arbitrary functions, or integrals along
both characteristics.  In the present paper we accept the later property as
definition.  As far as we know it was offered as independent definition in
the paper \cite{ZhISh} and the corresponding subclass was called 'Liouville
type equations'.  Since this term is suitable rather to join (or
intersection) of all subclasses mentioned above, we would say just 'equations
possessing integrals'.  Other term, 'Darboux integrable equations'
\cite{AK,ZhSS,SZh} is not exact, since Darboux had used rather the first of
the listed properties.

The given definition means that the lattice (\ref{uix}) admits the
$x$-integral $X$ and $i$-integral $I,$ that is functions on $x$ and finite
set of dynamical variables $\{u_{i+m},u^{(n)}_i=\pa^n_x(u_i)|\ m\in\Z,\
n\in\Z_+\}$ ($i$ is assumed to be fixed here), such that the identities hold
in virtue of the lattice
\[
    D_x(X)=0, \quad (T_i-1)(I)=0.
\]
The sequence of the coupled Riccati equations
\be                                                         \label{eg1}
    u_{i+1,x} - u_{i,x} = u^2_{i+1} - u^2_i
\ee
can be considered as the classical example, its integrals are of the form
\[
    I = u_{i,x}-u^2_i, \quad
    X = (u_i,u_{i+1},u_{i+2},u_{i+3})
      = {(u_i-u_{i+2})(u_{i+1}-u_{i+3})\over(u_i-u_{i+3})(u_{i+1}-u_{i+2})}.
\]
Slightly more complicated example is given by the lattice
\be                                                         \label{eg2}
    u_{i+1,x}u_{i,x} = u_{i+1}+u_i
\ee
for which, as one can easily prove,
\[
    I=(u_{i,xx}-1)^2/u^2_{i,x}, \quad
    X={(u_{i+3}-u_{i+1})(u_{i+2}-u_i)\over u_{i+2}+u_{i+1}}.
\]
It is not difficult to prove that in general case $I$ depends only on
derivatives $u^{(n)}_i,$ while $X$ depends only on shifts $u_{i+m}.$  We will
call the order of the higher derivative on which the $i$-integral depends the
order of $i$-integral, and the order of $x$-integral is equal to the
difference between the maximal and minimal shifts of the variable $u_i$ which
are involved in it.  So, in the last example $\ord I=2,$ $\ord X=3.$

To solve the lattice possessing integrals one has to solve the difference
equation $X=c_i$ where $c_i$ is an arbitrary sequence, and consistent with it
ordinary differential equation $I=C(x)$ where $C(x)$ is an arbitrary
function.  Usually this allows to obtain the explicit solution as the final
result, for example, one has for the lattice (\ref{eg1})
\[
    u_i= {\psi_{xx}\over2\psi_x}+{\psi_x\over z_i-\psi}
\]
where function $\psi(x)$ and parameters $z_i$ are arbitrary.  Notice, that
\[
    2I=S_\psi={\psi_{xxx}\over\psi_x}-{3\psi^2_{xx}\over2\psi^2_x}, \quad
    X=(z_i,z_{i+1},z_{i+2},z_{i+3}).
\]
Nevertheless, the relation between existence of integrals and explicit
formula for solution remains unclear till now even in the most studied
continuous case.

For the discrete equation (\ref{uij}) our definition of integrability means
existence of $i$- and $j$-integrals satisfying relations
\[
    (T_i-1)(I)=0, \quad (T_j-1)(J)=0.
\]
Now the dynamical variables are $u_{i+m,j}$, $u_{i,j+n}$, $m,n\in\Z$, since
the rest can be eliminated in virtue of (\ref{uij}).  It is easy to prove
that $i$-integral depends on the shifts $u_{i,j+n}$ only, and $j$-integral
depends on the shifts $u_{i+m,j}.$  The order of integrals is defined in just
the same way as order of $x$-integral for the equation (\ref{uix}).  As an
example, let us consider the equation
\be                                                         \label{eg3}
    u_{i+1,j+1} - u_{i,j+1} + {1\over u_{i+1,j}} - {1\over u_{ij}} = 0
\ee
which admits $i$-integral of the first order and $j$-integral of the third
order:
\[
    I = u_{i,j+1}+{1\over u_{ij}}, \quad
    J = {(u_{i+3,j}-u_{i+2,j})(u_{i+1,j}-u_{ij})\over
         (u_{i+3,j}-u_{i+1,j})(u_{i+2,j}-u_{ij})}.
\]

Now we summarize our results.  In the Section 2 the notion of the Laplace
invariants is transferred to the discrete case (cf. \cite{DN}).  The
sequences of Laplace invariants for the above examples and discrete Liouville
equations from the Section 3 are terminated by zero and this allows us to
conjecture, by analogy with continuous case, that this property is
characteristic for the equations possessing integrals \cite{AK} --
\cite{JuA}.  The main result of our paper is the proof of this conjecture in
one direction (necessity), given in the Section 5.  The outline of the proof
is the same as in continuous case.  In Section 6 the constructive formulae
are presented for the symmetries of the lattices (\ref{uix}) possessing
integrals.

Laplace invariants satisfy some recurrent relations analogous to the discrete
Toda lattice \cite{DN} -- \cite{Kor}.  In Section 4 these relations are
identified as the well known formulae for the B\"acklund transformations and
nonlinear superposition principle for the two-dimensional Toda lattice.  The
proof of sufficiency in the continuous case is based essentially on the
results on the structure of the integrals of this lattice \cite{Exp,Sh} and
therefore it would be very important to obtain analogous results in the
discrete case as well.

\section{Discrete Laplace invariants}

Laplace invariants of the linearized equation are very useful tool in the
theory of the equations (\ref{uxy}).  In particular, it was proved that the
equation possesses integrals iff the sequence of the Laplace invariants $h_j$
is terminated by zero at the both ends \cite{AK} -- \cite{JuA}.  The
invariants $h_{-1},h_0$ are calculated directly by the right hand side of the
equation and the rest are found recurrently by the formula
\be                                                         \label{hjxy}
    (\log h_j)_{xy} = h_{j+1}-2h_j+h_{j-1}
\ee
which is equivalent to the two-dimensional Toda lattice.

It turns out that analogous constructions can be defined in the discrete
situation as well.  Firstly let us consider differential-difference operator
\bean
    L = T_iD_x+a_iD_x+b_iT_i+c_i &=& (T_i+a_i)(D_x+b_{i-1})+H_{i,0} = \\
                                 &=& (D_x+b_i)(T_i+a_i)+K_{i,0}
\eean
where
\[
    H_{i,0}=c_i-a_ib_{i-1},\quad K_{i,0}=c_i-a_ib_i-D_x(a_i).
\]
One can easily check that if $H_{i,0}\ne0$ then the formula holds
\be                                                         \label{prixx}
    (D_x+b_{i,1})L = L_1(D_x+b_{i-1})
\ee
where
\[
    L_1 = (D_x+b_{i,1})(T_i+a_i) + H_{i,0}, \quad
    b_{i,1} = b_{i-1} - D_x(\log H_{i,0}).
\]
It means that the operator $D_x+b_{i-1}$ maps the kernel of $L$ into the
kernel of the operator $L_1.$ Moreover, operator ${1\over H_{i,0}}(T_i+a_i)$
define the inverse transformation which maps the kernel of $L_1$ into the
kernel of $L:$
\[
    L{1\over H_{i,0}}(T_i+a_i)=(T_i+a_i){1\over H_{i,0}}L_1.
\]
We shall call the mapping $L=L_0\mapsto L_1$ the Laplace $x$-transformation.
Its iterations yield the sequence of operators
\be                                                         \label{L+j}
    L_j = (T_i+a_i)(D_x+b_{i-1,j})+H_{ij},\quad j\ge0
\ee
where $H_{ij}$, $b_{ij}$ are defined by recurrent formulae
\bea
 & H_{i,j+1}=H_{ij}+D_x(a_i)+a_i(b_{i,j+1}-b_{i-1,j+1}),  & \nonumber\\
 & b_{i,j+1}=b_{i-1,j}-D_x(\log H_{ij}),\quad b_{i,0}=b_i. & \label{Hb}
\eea

Laplace $i$-transformation is defined in the similar manner.  Analog of the
formula (\ref{prixx}) looks
\be                                                  \label{prixi}
    (T_i + a_{i,1})L=L_{-1}(T_i+a_i)
\ee
where
\[
    L_{-1}=(T_i+a_{i,1})(D_x+b_i)+K_{i+1,0},\quad
    a_{i,1}=a_iK_{i+1,0}/K_{i,0}
\]
and it is assumed that $K_{i,0}\ne0.$  Operator ${1\over K_{i,0}}(D_x+b_i)$
defines the inverse transformation:
\[
    L{1\over K_{i,0}}(D_x+b_i)=(D_x+b_i){1\over K_{i+1,0}}L_{-1}.
\]
Iteration of the $i$-transformation generates the sequence of the operators
\be                                                         \label{L-j}
    L_{-j}=(D_x+b_{i+j})(T_i+a_{ij})+K_{ij}, \quad j\ge0
\ee
where $K_{ij}$, $a_{ij}$ are defined by recurrent relations
\bea
 & K_{i,j+1}= K_{i+1,j}-D_x(a_{i,j+1})
               -a_{i,j+1}(b_{i+j+1}-b_{i+j}), &             \nonumber \\
 & a_{i,j+1}=a_{ij}K_{i+1,j}/K_{ij},\quad a_{i,0}=a_i. &    \label{Ka}
\eea

It is more convenient for calculations to consider, under assumption
$a_i\ne0$ (this guarantees $a_{ij}\ne0$), the values
\[
    h_{ij}=\left\{\ba{cl}
              H_{ij}/a_i,           & j\ge0, \\
              K_{i,-j-1}/a_{i,-j-1},& j<0
           \ea\right.
\]
which we will call Laplace invariants.  Actually, this exchange is necessary,
since the operators $L$ should be considered up to the gauge $\tilde
L=\alpha^{-1}_{i+1}L\alpha_i$ and the values $h_{ij}$ are just the invariants
of this transformation.  In particular, comparison of the formulae
(\ref{prixx}), (\ref{prixi}) and their inverses demonstrates that the Laplace
$x$- and $i$-transformations are inverse to each other up to the gauge
transformation.  One can easily check that $h_{ij}$ satisfy the lattice
\be                                                         \label{hijx}
    \left(\log {h_{ij}\over h_{i+1,j}}\right)_x =
       h_{i,j-1}-h_{ij}-h_{i+1,j}+h_{i+1,j+1}
\ee
while the values $H_{ij},$ $K_{ij}$ do not satisfy closed system of
equations.

Laplace invariants of the lattice (\ref{uix}) are understood as Laplace
invariants of the linearization operator
\[
    L = T_iD_x - f_{u_{i,x}}D_x - f_{u_{i+1}}T_i - f_{u_i}.
\]
Passage from the values $H_{ij}$ and $K_{ij}$ to $h_{ij}$ is always possible
in virtue of the nondegeneracy condition $f_{u_{i,x}}\ne0.$  Notice also,
that the property $h_{i+1,j}=T_i(h_{ij})$ holds due to the shift-invariance
of the lattice and therefore the Laplace invariants can become zero only on
the whole line $j=j_0.$

Let us return to the examples presented in Introduction.  One can easily
check by direct calculation that for the lattice (\ref{eg1}) the invariants
$h_{i,-3}$ and $h_{i,0}$ become zero, as well as invariants $h_{i,-3}$ and
$h_{i,1}$ for the lattice (\ref{eg2}).  So the conjecture arises that the
criterion of the existence of the integrals mentioned above is valid also in
the discrete case.

Laplace transformations and invariants for the difference-difference
operators are defined quite similarly.  Moreover they were studied in the
review \cite{DN} and we only bring few key formulae.  Let us consider the
operator
\bean
    L = T_iT_j+a_{ij}T_j+b_{ij}T_i+c_{ij}
     &=& (T_i+a_{ij})(T_j+b_{i-1,j})+H_{ij,0} = \\
     &=& (T_j+b_{ij})(T_i+a_{i,j-1})+K_{ij,0}
\eean
where
\[
    H_{ij,0}= c_{ij} - a_{ij}b_{i-1,j}, \quad
    K_{ij,0}= c_{ij} - a_{i,j-1}b_{ij}.
\]
Operator $T_j+b_{i-1,j}$ maps the kernel of $L$ into the kernel of the
operator
\[
    L_1=(T_j+b_{ij,1})(T_i+a_{ij})+H_{i,j+1,0},
    \quad b_{ij,1}=b_{i-1,j}H_{i,j+1,0}/H_{ij,0}
\]
and operator ${1\over H_{ij,0}}(T_i+a_{ij})$ defines inverse transformation.
Iterations of this transformation, which we will call Laplace
$j$-transformation, yield the sequence of operators
\[
    L_k= (T_i+a_{i,j+k})(T_j+b_{i-1,jk}) + H_{ijk}, \quad k\ge0
\]
where $H_{ijk}$, $b_{ijk}$ are defined recurrently:
\[
    H_{ij,k+1}=H_{i,j+1,k}+a_{i,j+k}b_{ij,k+1}-a_{i,j+k+1}b_{i-1,j,k+1},
\]\[
    b_{ij,k+1}=b_{i-1,jk}H_{i,j+1,k}/H_{ijk},\quad b_{ij,0}=b_{ij}.
\]
Laplace $i$-transformation and values $K_{ijk},$ $a_{ijk}$ are defined
analogously.

When $a_{ij}b_{ij}\ne0$ one can introduce Laplace invariants
$h_{ijk}=H_{ijk}/(a_{i,j+k}b_{i-1,jk})$ for $k\ge0,$
$h_{ijk}=K_{ij,-k-1}/(a_{i,j-1,-k-1}b_{i-k-1,j})$ for $k<0.$  As in the
previous case these quantities do not change under the gauge transformations
and satisfy closed system of equation
\be                                                         \label{hijk}
    (h_{i,j+1,k-1}+1)(h_{i+1,j,k+1}+1) =
     (h^{-1}_{i,j+1,k}+1)(h^{-1}_{i+1,j,k}+1)h_{ijk}h_{i+1,j+1,k}.
\ee
It is easy to check that for the example (\ref{eg3}) invariants $h_{ij,-3}$
and $h_{ij,0}$ of the linearized equation become zero in accordance with our
conjecture.

\section{Discrete Liouville equations}

Remind that the Liouville equation \cite{Lio} can be obtained from the Toda
lattice (\ref{hjxy}) by imposing the simplest boundary conditions
$h_1=h_{-1}=0.$  The natural question arise, what is the result of the same
reduction of the discrete Toda lattices (\ref{hijx}) and (\ref{hijk}).  We
will see that structure and properties of these three equations are very
similar: each of them possesses second order integrals, has only two nonzero
Laplace invariants, and admits linearizing substitutions of the first order.
Moreover, the passage to the limit turns the discrete versions into the
Liouville equation itself.  Let us consider these examples in more details.

Boundary conditions $h_{i,1}=h_{i,-1}=0$ turn the equation (\ref{hijx}) into
the lattice of the form (\ref{uix}) on the variables $u_i=h_{i,0}:$
\be                                                         \label{Li2}
    u_{i+1,x}u_i-u_{i+1}u_{i,x} = u_{i+1}u_i(u_{i+1}+u_i).
\ee
It is easy to check that assuming $u_i(x)=\eps u(x,y),$ $y=\eps i$ and
passing to the limit $\eps\to0$ one obtains the Liouville equation $(\log
u)_{xy}=2u.$  Notice also that the lattice (\ref{Li2}) can be obtained from
(\ref{eg1}) by means of substitution $\tilde u_i=u_{i+1}-u_i.$

The linearization operator for (\ref{Li2}) is
\[
    L = D_xT_i - {u_{i+1}\over u_i}D_x - (u_{i,x}/u_i+2u_{i+1}+u_i)T_i
               + u_{i+1}(u_{i,x}/u^2_i-1)
\]
and calculating of the Laplace invariants gives $h_{i,1}=h_{i,-2}=0.$

Separating the variables $u_i$ and $u_{i+1}$ and differentiating yields
\[
    {u_{i+1,x}\over u_{i+1}} - u_{i+1} = {u_{i,x}\over u_i} + u_i, \quad
    {u_{i+1,xx}\over u_{i+1}}-{u^2_{i+1,x}\over u^2_{i+1}} - u_{i+1,x} =
    {u_{i,xx}\over u_i}-{u^2_{i,x}\over u^2_i} + u_{i,x}
\]
from which the $i$-integral can be easily found:
\[
    I = 2{u_{i,xx}\over u_i}-3{u^2_{i,x}\over u^2_i}-u^2_i.
\]
It is not difficult to find the $x$-integral as well:
\[
    X = (1+u_i/u_{i+1})(1+u_i/u_{i-1}).
\]
Equation (\ref{Li2}) is obtained from the linear equation
$v_{i+1,x}-v_{i,x}=0$ by means of the substitution
\[
    u_i={(v_{i+1}-v_i)v_{i,x}\over v_{i+1}v_i}
\]
(cf. substitution $u=v_xv_y/v^2$ which linearize the Liouville equation) and
this allows at once to write down the formula for the general solution:
\[
    u_i(x) = {(c_{i+1}-c_i)\psi_x\over(c_{i+1}+\psi)(c_i+\psi)}
\]
where $c_i$ are arbitrary constants and $\psi(x)$ is arbitrary function.

In the totally discrete case the boundary conditions $h_{ij,1}=h_{ij,-1}=0$
reduce the equation (\ref{hijk}) to equation of the form (\ref{uij}) on the
variables $u_{ij}=h_{ij,0}:$
\be                                                         \label{Li3}
    u_{i+1,j+1}(1+1/u_{i+1,j})(1+1/u_{i,j+1})u_{ij} = 1.
\ee
It turns into the lattice (\ref{Li2}) after the passage to limit
$u_{ij}=-\eps u_i(\eps j),$ $\eps\to0.$

The Laplace invariants of this equation $h_{ij,1}$ and $h_{ij,-2}$ are equal
to zero just as in the previous example.  The integrals are given by formulae
\[
    I= \left(1+{u_{ij}(u_{i,j-1}+1)\over u_{i,j-1}}\right)
       \left(1+{u_{ij}\over u_{i,j+1}(u_{ij}+1)}\right),
\]\[
    J= \left(1+{u_{ij}(u_{i-1,j}+1)\over u_{i-1,j}}\right)
       \left(1+{u_{ij}\over u_{i+1,j}(u_{ij}+1)}\right).
\]
The substitution
\[
    u_{ij}= -{(v_{i+1,j}-v_{ij})(v_{i,j+1}-v_{ij})\over v_{i+1,j}v_{i,j+1}}
\]
brings (\ref{Li3}) to the linear equation
$v_{i+1,j+1}-v_{i+1,j}-v_{i,j+1}+v_{ij}=0$ and therefore the formula for the
general solution is valid
\[
    u_{ij}= -{(c_{i+1}-c_i)(k_{j+1}-k_j)\over(c_{i+1}+k_j)(c_i+k_{j+1})},
\]
where $c_i,k_j$ are arbitrary constants.

\section{Discrete Toda lattices}

Now we slightly digress from our main subject in order to discuss the
equations (\ref{hijx}) and (\ref{hijk}) which are the natural discrete
analogues of the Toda lattice (\ref{hjxy}).  These equations have been
already studied in literature from this point of view (possibly in somewhat
different notation) see e.g. \cite{DN} -- \cite{Kor}.  We should like to
reproduce for completeness some well-known formulae which demonstrate that
the links between these three lattices are more close than it seems at the
first glance.  The results of this Section will not be necessary in what
follows.

It is more convenient to start from the lattice
\be                                                         \label{rjxy}
    r_{j,xy} = \exp(r_{j+1}-2r_j+r_{j-1}),
\ee
rather than from (\ref{hjxy}) which is related with it by means of
substitution $h_j=r_{j,xy}.$  The B\"acklund transformation for this lattice
is of the form
\bea                                                        \label{rijx}
 (r_{ij}-r_{i+1,j})_x    &=& \exp(r_{i,j-1}-r_{ij}-r_{i+1,j}+r_{i+1,j+1}),\\
                                                            \label{rijy}
 (r_{i+1,j}-r_{i,j-1})_y &=& \exp(r_{ij}-r_{i+1,j}-r_{i,j-1}+r_{i+1,j-1}).
\eea
Indeed, the cross-differentiation yields the formula
\[
    r_{ij,xy} = \exp(r_{i,j+1}-2r_{ij}+r_{i,j-1})+c(x,y)
\]
where function $c(x,y)$ does not depend on $i,j$ and can be easily removed by
shift of the form $r\to r+C(x,y).$  The lattice (\ref{rijx}) is reduced to
(\ref{hijx}) by means of substitution $h_{ij}=(r_{ij}-r_{i+1,j})_x$.

We see that the second subscript in the formulae (\ref{rijx}), (\ref{rijy})
corresponds to the shift in the lattice (\ref{rjxy}) while the first one
denotes the iteration of the B\"acklund transformation.  Now introduce the
third subscript and consider the B\"acklund transformation which changes it
and leaves the first subscript fixed (vice versa, the fixed third subscript
must be assigned to the variable $r$ in the formulae (\ref{rijx}),
(\ref{rijy})):
\bean
   (r_{ijk}-r_{ij,k+1})_x     &=&
     \exp(r_{i,j-1,k}-r_{ijk}-r_{ij,k+1}+r_{i,j+1,k+1}), \\
   (r_{ij,k+1}-r_{i,j-1,k})_y &=&
     \exp(r_{ijk}-r_{ij,k+1}-r_{i,j-1,k}+r_{i,j-1,k+1}).
\eean
Let us consider the superposition of these B\"acklund transformations.  It is
sufficient to consider only $x$-part, since differentiating on $y$ brings to
the same result.  Calculating $(r_{ijk}-r_{i+1,j,k+1})_x$ in two different
ways one obtains after some transformations the relation
\bean
&& (T_j-1)(\exp(r_{i+1,j-1,k}-r_{i+1,j,k+1}-r_{i,j-1,k}+r_{ij,k+1}) - \\
&& \qquad\qquad -\exp(r_{i,j-1,k+1}-r_{i+1,j,k+1}-r_{i,j-1,k}+r_{i+1,jk}))=0.
\eean
The expression in the brackets is equal to some constant $c_{ik}$ which can
be set to 1 by shift of the form $r_{ijk}\to r_{ijk}+C_{ik},$ resulting in
desired difference equation
\bea
   &  \exp(r_{i+1,j-1,k}+r_{ij,k+1})
     -\exp(r_{i,j-1,k+1}+r_{i+1,jk}) = & \nonumber \\
   & = \exp(r_{i+1,j,k+1}+r_{i,j-1,k}).& \label{rijk}
\eea
It is equivalent, up to the linear transformation of the coordinate axes, to
the well-known Hirota-Miwa equation \cite{H81,Miwa}
\[
    af_{i+1,jk}f_{i-1,jk}+bf_{i,j+1,k}f_{i,j-1,k}+cf_{ij,k+1}f_{ij,k-1}=0
\]
(it is clear that all coefficients can be removed by means of scaling
$f\to\alpha^{i(i-1)}\beta^{j(j-1)}f$).  Substitution
\[
    h_{ijk}=\exp(r_{i+1,j,k+1}-r_{i,j-1,k+1}-r_{i+1,jk}+r_{i,j-1,k})
\]
maps equation (\ref{rijk}) into equation
\[
    (h_{i+1,j,k+1}+1)(h_{i,j-1,k}+1) =
     (h^{-1}_{i,j-1,k+1}+1)(h^{-1}_{i+1,jk}+1)h_{i+1,j-1,k}h_{ij,k+1}
\]
which coincides with (\ref{hijk}) up to the linear change of $i,j,k.$

It should be stressed that the changes from the variable $r$ to the variable
$h$ are different in all three cases, and therefore one cannot say that the
lattice (\ref{hijx}) defines the B\"acklund transformation for (\ref{hjxy})
and the equation (\ref{hijk}) is its nonlinear superposition principle.
Nevertheless, we have demonstrated that these three equations live on the
same three-dimensional lattice generated by the B\"acklund transformations of
the lattice (\ref{rjxy}).

\section{Necessary conditions for existing of integrals}

In this Section we prove the main theorem which allows to check, rather
effectively, if the given equation (\ref{uix}) or (\ref{uij}) possesses the
integrals.

We consider in details only the lattices (\ref{uix}).  Let $g$ be function on
$x$ and dynamical variables $u_{i+m},$ $u^{(n)}_i.$  Denote $g_*$ its
linearization, that is operator
\[
    g_*=  \sum^\infty_{m=-\infty} g_{u_{i+m}}T^m_i
         +\sum_{n=1}^\infty g_{u^{(n)}_i}D^n_x.
\]
Intermediate calculation proves that the properties hold
\be                                                         \label{DT*}
    (D_x(g))_*=D_xg_* \ \bmod [L], \quad
    (T_i(g))_*=T_ig_* \ \bmod [L]
\ee
where $L$ is linearization operator of the lattice (\ref{uix}) and $[L]$
denotes set of all operators of the form $ML$ with arbitrary
differential-difference operators $M.$

\begin{theorem}
If the lattice (\ref{uix}) possesses the $x$-integral $X$ of the order $n$
then its Laplace invariant $h_{i,-j}$ vanishes for some $j,$ $0<j\le n.$
Analogously, if it admits the $i$-integral $I$ of the order $n$ then the
invariant $h_{ij}$ vanishes for some $j,$ $0\le j<n.$
\end{theorem}
\proof
Let us prove by contradiction the first statement.  Assume that
$h_{i,-j}\ne0$ for all $j=1,\dots,n.$  It is the same as the quantities
$K_{ij}\ne0$ for all $j=0,\dots,n-1$ and therefore the formulae (\ref{L-j}),
(\ref{Ka}) define the operators $L_{-j}$ for $j=0,\dots,n.$  The relation
(\ref{prixi}) for these operators takes form
\[
    (T_i+a_{i,j})L_{-j+1}=L_{-j}(T_i+a_{i,j-1}), \quad j=1,\dots,n.
\]
The repeated application of this formula brings to $L_{-j}A_{j-1}=0\ \bmod
[L]$ where the difference operators $A_j$ are defined as follows
\[
    A_{-1}=1, \quad A_j=(T_i+a_{ij})A_{j-1}, \quad j=0,\dots,n.
\]
Therefore one obtains the relation
\[
    D_xA_j= -b_{i+j}A_j-K_{ij}A_{j-1}\ \bmod [L], \quad j=1,\dots,n.
\]

Due to the shift $T_i$ one can assume without loss of generality that $X$
depends on $u_i,\dots,u_{i+n}.$  Consider the expansion of $X_*$ over the
operators $A_j:$ $X_*=\sum^n_{j=0}\xi_jA_{j-1}$ and multiply it by $D_x$ from
the left.  Collecting the coefficients on $A_j$ and $D_x$ in the relation
$D_xX_*=0\ \bmod [L]$ brings to the sequence of equations
\bean
    (D_x-b_{i+n-1})(\xi_n) &=& 0,\\
    (D_x-b_{i+j-1})(\xi_j) &=& \xi_{j+1}K_{ij},\quad j=1,\dots,n-1, \\
                D_x(\xi_0) &=& \xi_1K_{i,0},\\
                   \xi_0 &=& 0
\eean
from which the contradiction follows: $\xi_0=\dots=\xi_n=0\ \Rightarrow\
X_*=0.$

The proof of the second statement is quite analogous and we just outline it.
Assume that $H_{ij}\ne0$ for all $0\le j<n$ and define the operators $L_j$
accordingly to (\ref{L+j}), (\ref{Hb}) and operators $B_j$ by formulae
\[
    B_{-1}=1, \quad B_j=(D_x+b_{i-1,j})B_{j-1}.
\]
Applying the relations
\[
    T_iB_j= -a_iB_j-H_{ij}B_{j-1}\ \bmod [L],\quad j=1,\dots,n
\]
to the equation $T_iI_*=I_*\ \bmod [L]$ where $I_*=\sum^n_{j=0}\xi_jB_{j-1}$
one obtains the sequence of equations
\bean
    (a_iT_i+1)(\xi_n) &=& 0,\\
    (a_iT_i+1)(\xi_j) &=& -\xi_{j+1}H_{ij},\quad j=1,\dots,n-1, \\
                \xi_0 &=& -\xi_1H_{i,0},\\
           T_i(\xi_0) &=& 0
\eean
which implies contradiction $I_*=0.$
\qed
\medskip

In the totally discrete case the following statement can be proved in similar
manner.

\begin{theorem}
Let equation (\ref{uij}) possesses the $i$-integral ($j$-integral) of the
order $n.$  Then the Laplace invariant $h_{ijk}=0$ for some $k,$ $0\le k<n$
($h_{ij,-k}=0$, $0<k\le n$).
\end{theorem}

\section{Higher symmetries}

The notion of symmetry for the equations (\ref{uix}), (\ref{uij}) and
(\ref{uxy}) is introduced uniformly as any function on the dynamical
variables of the given equation which satisfies the characteristic equation
$L(g)=0$ where $L$ is linearization operator.  Less formally, symmetry can be
understand as equation $u_t=g$ compatible with the given one.  It is known
\cite{ZhSS} that equations (\ref{uxy}) possessing integrals admit also the
rich family of the symmetries of the form
\[
    g= P(Y)+Q(X)
\]
where $P$ and $Q$ some explicitly computable differential operators on $D_x$
and $D_y$ respectively and $X,$ $Y$ are arbitrary $x$- and $y$-integrals.

The aim of this Section is to obtain the analogous formula in the discrete
case.  Let us prove the following Lemma as a preliminary.

\begin{lemma}
Let the lattice (\ref{uix}) possesses integrals then the kernels of operators
$D_x-f_{u_{i+1}},$ $T_i-f_{u_{i,x}}$ are not empty in the space of the
functions on dynamical variables.  Analogously, if the equation (\ref{uij})
possesses integrals then the kernels of the operators $T_i-f_{u_{i,j+1}},$
$T_j-f_{i+1,j}$ are not empty.
\end{lemma}
\proof
It is sufficient to differentiate the relation defining the integral on the
higher dynamical variable.  For example, differentiating on $u^{(n)}_i$ of
the relation $(T_i-1)I(x,u_i,\dots,u^{(n)}_i)=0$ where
$u^{(j+1)}_{i+1}=D^j_x(f)$ in virtue of the lattice (\ref{uix}) yields
$(f_{u_{i,x}}T_i-1)(I_{u^{(n)}_i})=0,$ therefore $1/I_{u^{(n)}_i}$ belongs to
the kernel of the operator $T_i-f_{u_{i,x}}.$
\qed
\medskip

Now let us prove the theorem about the symmetries of the lattice (\ref{uix})
(we restrict ourselves by this case since the arguments for the equation
(\ref{uij}) are quite analogous).

\begin{theorem}
Let the lattice (\ref{uix}) admits an $x$-integral of the order $m$ and an
$i$-integral of the order $n.$  Then the differential operator $P$ and the
difference operator $Q$ exist of orders not greater than $m-1$ and $n-1$
respectively such that the formula
\[
    g = P(I)+Q(X)
\]
defines the symmetry of the given lattice for any $i$-integral $I$ and
$x$-integral $X.$
\end{theorem}
\proof
Accordingly to the Theorem 1, the existence of the $x$-integral implies that
$h_{i,-j}=0$ for some $j,$ $0<j\le m.$  Applying the Laplace
$i$-transformation (\ref{L-j}), (\ref{Ka}) for $j-1$ times one obtains the
general solution of the equation $L(g)=0$ in the form
\[
    g = {1\over K_{i,0}}(D_x+b_i)\cdot\dots\cdot
        {1\over K_{i,j-2}}(D_x+b_{i+j-2})(\psi)
\]
where $\psi$ satisfies equation
\[
    (D_x+b_{i+j-1})(T_i+a_{i,j-1})(\psi)=0.
\]
Let $(T_i+a_{i,j-1})(\psi)=0,$ then $\psi$ can be represented as
$\psi=K_{i,j-2}\cdots K_{i,0}\phi I$ where $I$ is an arbitrary $i$-integral
and $\phi$ satisfies equation $(T_i+a_i)(\phi)=0.$  Solvability of this
equation was proved in Lemma 3 (notice that it is provided by $i$-integral)
resulting in particular solution of the form $g=P(I)$ of the characteristic
equation.

Now consider Laplace $x$-transformation.  Let the invariant $h_{ij}$ vanish,
$0\le j<n,$ then the solution of the characteristic equation can be
represented as
\[
    g = {1\over H_{i,0}}(T_i+a_i)\cdot\dots\cdot
        {1\over H_{i,j-1}}(T_i+a_i)(\psi)
\]
where $\psi$ satisfies equation
\[
    (T_i+a_i)(D_x+b_{i-1,j})(\psi)=0.
\]
Moreover, accordingly to (\ref{Hb}),
\[
    b_{i-1,j} = b_{i-j-1,0}-D_x\log(H_{i-1,j-1}\dots H_{i-j,0})
\]
and therefore the function $\psi=H_{i-1,j-1}\dots H_{i-j,0}\phi X$ is the
particular solution if $X$ is arbitrary $x$-integral and $\phi$ satisfies
equation $\phi_x+b_{i-j-1,0}\phi=0.$  This equation is solvable according to
Lemma 3, and one obtains the particular solution of the determining equation
of the form $g=Q(X).$
\qed
\medskip

As an example we present the formula for the symmetries of the discrete
Liouville equation (\ref{Li2}):
\[
    u_{i,t} = D_x(u_iI) + (T_i-1)((u_i+{u^2_i\over u_{i+1}})X).
\]
The first term in this formula describes the family of the evolution
equations for which the lattice (\ref{Li2}) defines the B\"acklund
transformation.  Note for comparison that in the continuous case operators
$P$ and $Q$ for the symmetries of the equation $(\log u)_{xy}=2u$ are
$P=D_xu,$ $Q=D_yu.$

\bigskip

Authors are grateful to A.V.~Zhiber, V.V.~Sokolov, I.T.~Habibullin,
S.P.~Tsarev and A.B.~Shabat for interest to the work and useful discussions.
The work is supported by Russian Foundation for Basic Researches.


\end{document}